\documentclass[twocolumn,amsmath,amssymb,pre]{revtex4-1}
\usepackage{graphicx}
\usepackage{dcolumn}
\usepackage{overpic}
\usepackage{float} 
\usepackage{bbm}

\usepackage{setspace}

\begin{document}

\title{Tuning the Aharonov-Bohm effect with dephasing in nonequilibrium transport}

\author{ Georg Engelhardt$^1$    }

\author{ Jianshu Cao$^{1,2} $}
\email{jianshu@mit.edu}

\affiliation{%
$^1$Beijing Computational Science Research Center, Beijing 100193, Peopleʼs Republic of China\\
$^2$Department of Chemistry, Massachusetts Institute of Technology, 77 Massachusetts Avenue,
Cambridge, Massachusetts 02139, USA
}

\date{\today}


\begin{abstract}
The Aharanov-Bohm (AB) effect, which predicts that a magnetic field strongly influences the wave function of an electrically charged particle, is investigated in a three site system  in terms of the quantum control by an additional dephasing source. The AB effect leads to a non-monotonic dependence of the steady-state current on the gauge phase associated with the molecular ring. This dependence is sensitive to site energy, temperature, and dephasing, and can be explained using the concept of the dark state. Although  the phase effect vanishes in the steady-state current for strong dephasing, the phase dependence remains visible in an associated waiting-time distribution, especially at short times. Interestingly, the phase rigidity (i.e., the symmetry of the AB phase) observed in the steady-state current is now broken in the waiting-time statistics, which can be explained by the interference between transfer pathways.
\end{abstract}
\maketitle

\section{Introduction.}
The celebrated Aharonv-Bohm (AB) effect predicts that an electromagnetic potential generated by a magnetic field influences the complex-valued wave function of  electrically charged particles~\cite{Bohm2003}. This  genuine quantum effect with no classical counterpart has been verified in interference experiments~\cite{Chambers1960,Osakabe1986,VanOudenaarden1998,Webb1985,Bachtold1999,Aharony2014}.
Moreover, the relation of spin-orbit coupling and the AB effect has been investigated and discussed in relation to thermal properties~\cite{Aharony2011,Tu2012,Liu2016}. 
Yet,  the coherent dynamics of quantum particles is  sensitive to dephasing processes, which appear due to  couplings to the thermal environment. For this reason,  signatures of the AB effect will disappear for a strong system-environment coupling strength.

\begin{figure}[t]
\includegraphics[width=1\linewidth]{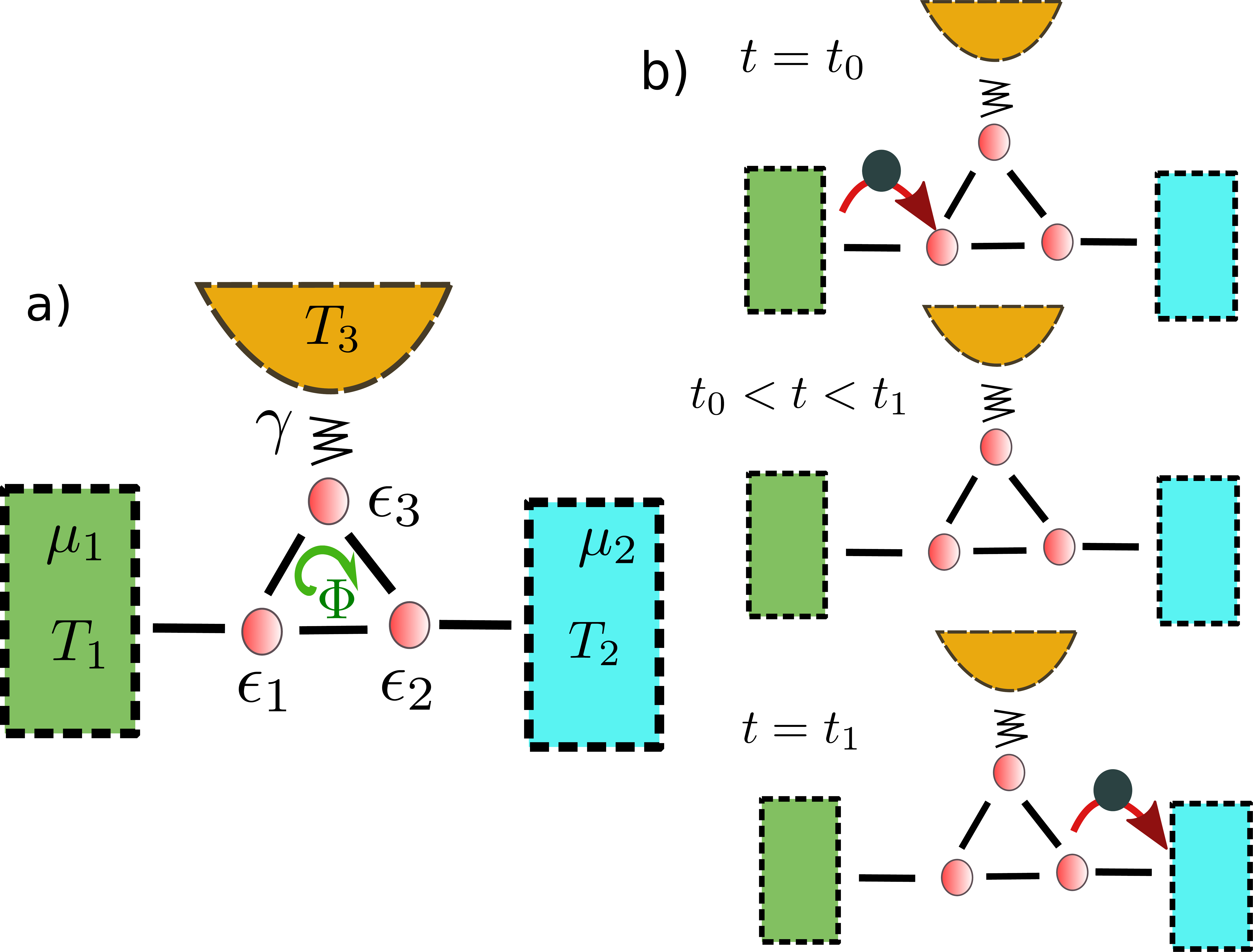}
\caption{ (a) A molecular junction consisting of three sites $n=1,2,3$ which are connected to two electronic leads $n=1,2$ and a thermal bath $n=3$. (b) Sketch of the waiting time experiment. A particle jumping into the system at $t=t_0$ is subjected to a coherent time evolution during $t_0<t<t_1$, before it jumps into reservoir $n=2$ at time $t=t_1$. The waiting time distribution describes the jump probability as a function of the time difference $\tau= t_1-t_0$.
}
\label{fig:overview}
\end{figure}%
%

%
To study the interplay of the AB effect and dephasing in detail,  we investigate  the transport characteristics of a minimal AB ring consisting of three sites, which is  coupled to a dephasing heat bath, and its potential for quantum manipulation and control. To explore the related basic questions and discuss fundamental aspects, we choose this  generic model system with potential applications to more realistic systems and devices.

 We are specially interested in the impact of the crossover  from weak to strong coupling on the AB effect. It is known from investigations of excitonic and electronic dynamics, that the interplay of coherent and incoherent dynamics can give rise to interesting consequences, such as the enhancement of transport efficiency~\cite{Wang2014}. To  calculate the dynamical properties of the system, we apply the \textit{poloran transformation}, which is frequently used in molecular systems~\cite{Silbey1984,Harris1985}. It allows for an accurate  theoretical treatment of arbitrary system-bath coupling strengths on equal footing. The application of the polaron transformation to transport problems such as the AB effect has been only recently studied.

For weak coupling (compared to the energies of the ring system), we find that the signatures of the AB effect in the steady-state transport are  well visible, although they are attenuated to some extend. For strong coupling, we find that the AB signature in the steady state nearly completely vanishes. In contrast, we find that the impact of the AB effect is  apparent in a waiting-time experiment even  for a very strong system-bath coupling.

In order to pronounce the consequences of the  AB effect on the current, we choose the system parameters so that the three-level system exhibits a \textit{dark state}.  A dark state is an eigenstate, at which the wave function on a local site vanishes  due to a completely destructive interference of two coherent paths which a particle can take. This effect has been intensively investigated  in quantum optics~\cite{Scully1992,Scully1991}. Besides, there are also investigations about the role of dark states in electron transport~\cite{Poeltl2013,Emary2007}. In particular, in combination with the Coulomb blockade, this effect can be used for a quantum control of the transport properties. Our model and methods are well suited to demonstrate new aspects of dark states due to strong environmental coupling and show how to control the dark state mechanism in fermionic transport instead of optics.


Our general findings are relevant for various experimental systems. Electronic semiconductor nanostructures with their  high experimental control of system parameters and quantum states  allow for  precise measurements of the current and its related full counting statistics~\cite{Flindt2009,Wagner2017,Averin2011,Camati2016,Gaudreau2006,Rogge2008,Fujisawa2006}, and have been shown to exhibit the Aharnov-Bohm effect~\cite{Kobayashi2002,Strambini2009}. Single molecular junctions, which consist of a molecule bridging two electronic leads~\cite{Aradhya2013,Tao2006,Nitzan2003,Zimbovskay2013,Xiang2016,Jiang2015} and allow for high-precision current and fluctuation meashurements~\cite{Secker2011,Tal2008,Neel2011,Djukic2006,Tsutsui2010}, have shown clear signatures of quantum interference in molecular ring structures~\cite{Guedon2012,Vazquez2012}. The electronic dynamics on the molecules is often subjected to strong dephasing due to the coupling to vibrational modes and the environment, which can be effectively described with our methods. 
Furthermore, using time-dependent optical fields enables the creation of artificial gauge-field of uncharged particles in cold-atoms experiments~\cite{Aidelsburger2013,Jotzu2014}.

\section{The system}

As sketched in Fig.~\ref{fig:overview}(a), the system is coupled to  two  electronic reservoirs $n=1,2$ and contains three sites $n=1,2,3$, which are arranged in a ring geometry. This is a minimal model which is capable of exhibiting the celebrated AB effect. 
 The electronic dynamics in molecules is often strongly coupled to  the molecular environment. To take this into account, we  assume that site $n=3$ is additionally coupled to a  thermal bath.

\subsection{The Hamiltonian} 
 
The Hamiltonian describing the system reads  
\begin{equation}
H_{\rm th} = H_{\rm s} + H_{\rm s,b} +H_{\rm b}+ H_{\rm s,r} +H_{\rm r}+ H_{\text{int} },
\label{eq:hamiltonian}
\end{equation}
where
\begin{align}
H_{\rm s} & = \sum_{n=1}^{3} \epsilon_n \hat a_n^\dagger \hat a_n  + \sum_{n,m=1}^{3} J_{n,m}  \hat a_n^\dagger \hat a_m,  \label{eq:electronHamiltonian}\\
H_{\rm s,b} &= \sum_{k} V_{k}  \hat a_3^\dagger \hat a_3 \left(\hat b_{k} + \hat b_{k}^\dagger \right), \;
\; H_{\rm b} =\sum_{k}\hat \omega_{k} b_{k}^\dagger \hat b_{k}, \nonumber \\
H_{\rm s,r} &= \sum_{n=1}^{2}\sum_{k} W_{n,k} \left( \hat a_n^\dagger \hat c_n +  h.c. \right), \;
\; H_{\rm r} =\sum_{n=1}^{2} \sum_{k} \epsilon_{n,k} \hat c_{n,k}^\dagger \hat c_{n,k}.\nonumber 
\end{align}
The  system $H_{\rm s}$  is coupled to the electronic reservoirs $H_{\rm r}$  and heat bath  $H_{\rm b}$  via the Hamiltonians $H_{\rm s,r}$ and $H_{\rm s,b}$.
The coupling between system and electronic reservoirs $H_{\rm s,r}$ is assumed to be weak so that an application of the Redfield equation with Born-Markov approximation is valid, but we allow for a strong coupling between system and heat bath $H_{\rm s,b}$. The operators $\hat a_n$, $\hat c_{n,k}$ shall fulfill fermionic commutation relations,  while $\hat b_{n,k}$ are bosonic operators. The term $H_{\text{int} }$  describes the Coloumb interaction of two particles, which is assumed to be so strong that at most one particle can be in the system. For this reason, this term is not further specified.
 
The AB effect in a tight-binding model can only appear in a ring geometry. Let us consider a parameterization of the internal system coupling parameters $J_{n,m}=J_{n,m}^{(0)} e^{i \phi_{n,m}} $ with real $J_{n,m}^{(0)}$ and $\phi_{n,m}$. The coupling between  the reservoirs and the system is parameterized accordingly, namely $ W_{n,k}=W_{n,k}^{(0)} e^{i \varphi_{n,k}} $ with real $W_{n,k}^{(0)}$, $\varphi_{n,k}$. The phase factors $e^{i \varphi_{n,k}}$ can be transformed away by  gauge transformations $\hat c_{n,k}\rightarrow \hat c_{n,k}e^{-i \varphi_{n,k}}$, rendering  $W_{n,k}$ real valued. However, if all $J_{n,m}^{(0)}\neq 0$, it is not possible to completely render the coupling parameters $J_{n,m}$ real valued by gauge transformations. In contrast, the phase
\begin{equation}
\phi=\phi_{2,1}+\phi_{3,2}+\phi_{1,3}\mod 2\pi
\end{equation}
can be proven to be invariant under  transformations $\hat a_{n}\rightarrow \hat a_{n}e^{-i \tilde  \phi_{n}}$ for arbitrary $\tilde  \phi_{n}$. The phase $\phi$ is thus a gauge-invariant quantity and has physical relevance, i.e., the energies and the eigenstates of the Hamiltonian depend on $\phi$. This is the underlying reason for the famous AB effect. In contrast, if the ring is broken, i.e.,  $J_{n,m}^{(0)}=0$ for one coupling, the phases can be removed. For example, such a phase  $\phi$ appears if there is a magnetic field perpendicular to the ring~\cite{Bernevig2013}. More precisely, the phase can be expressed in terms of a magnetic field as $\phi = \frac {e}{h}\cdot B_{\rm n} \cdot A$, where $B_{\rm n}$ is the component of the magnetic field normal to the ring, $A$ is the ring area and $\frac {h}{2e}= 2,068 \cdot 10^{-15}\rm\, T m^2$ is the magnetic flux quantum.  For a notational reason, we define $J_n= J_{n,n+1}^{(0)}$, where $n=4$ corresponds to $n=1$.

\subsection{Dark state}

\label{sec:darkState}

A dark state denotes an eigenstate at which the wave function on one site of the local basis vanish completely due to destructive interference. Because of this outstanding property, dark states  have been suggested as quantum memory~\cite{Fleischhauer2002} and are related to electromagnetically-induced transparency~\cite{Boller1991,Lukin1997,Scully1992,Scully1991}.

If  two  onsite energies $\epsilon_{n}$ are equal, $J_n=J$ and $\phi=0,\pi$, then the system can be easily diagonalized due to the appearance of a dark state. Assuming, e.g., $\epsilon_1=\epsilon_3$, we find  the eigenstates
\begin{equation}
 \left| \Psi_{0,\pi}^{\text{dark}} \right> = \frac{1}{\sqrt{2}}\left( \hat a_{1}^\dagger \mp \hat a_{3}^\dagger \right) \left|\text{ vac} \right>.
\end{equation}
with energies 
$
  E_{0,\pi}^{\text{dark}}  = \epsilon_3 \mp J , 
$
for $\phi=0,\pi$, respectively.
The wave function is valid for the gauge choice of $\phi_{1,2}=\phi_{2,3}=0$ and $\phi_{3,1}=0,\pi$. As these states have no overlap with site $n=2$, they are indeed dark states, which have an essential influence on the transport properties.

\section{Methods}

\subsection{Polaron-transformed Redfield equation}

\label{sec:polaronTrafo}

To probe experimentally accessible quantities, we  focus on the steady-state electronic current of particles entering reservoir $n=2$ in the long-time limit of $t\rightarrow \infty$, the corresponding noise, and the waiting-time statistics. 

We take advantage of the polaron-transformed Redfield equation~\cite{Wang2015,Xu2016,Xu2016a}. 
The polaron transformation is defined by a unitary transformation
\begin{equation}
\hat U = e^{\hat S \hat a_{3}^\dagger \hat a_3 }  \qquad \text{with}\qquad  \hat S = \sum_{k} \frac{V_{k}  }{ \omega_k } \left(\hat b_{k}^\dagger - \hat b_{k}  \right) .
\label{eq:darkStateWF}
\end{equation}
Details can be found in Appendix~\ref{sec:polaronTransformation}.
The  polaron transformation modifies the Hamiltonian Eq.~\eqref{eq:electronHamiltonian} as follows:%
\begin{align}
J_{2} &\rightarrow    \kappa \cdot J_{2},\quad
J_{3} \rightarrow    \kappa \cdot J_{3},\quad
\epsilon_3\rightarrow \epsilon'_3 = \epsilon_3 - \sum_j \frac{\left| V_k\right| ^2}{\omega_k}
 \label{eq:renormalization}\\
H_{\rm s,b} &\rightarrow H_{\rm s,b}^{(p) } =\sum_{j=1,2}  J_{j} \hat a_{3}^\dagger \hat a_{j} \hat V_{3,j}+ \text{ h.c. },
\end{align} 
with
\begin{align}
	0< \kappa   &= e^{- \frac 12 \sum_{k} \frac{ V_{k}^2}{\omega_k  } \coth(\beta \omega_k/2)  }<1 , 
	\label{eq:kappa}\\
	\hat V_{3,j} &= e^{\hat S }- \kappa.
	\label{eq:couplingOperator}
\end{align}
Here, $\kappa$ is a renormalization parameter, which captures the influence of the strongly coupled bath on the system, i.e., the onsite energy of site $n=3$ gets renormalized. As explained in Sec.~\ref{sec:darkState}, the constrain $\epsilon_3=\epsilon_1$ is a requirement for the appearance of the dark state. Yet, due the coupling to the heat bath, this condition has to be fulfilled by $\epsilon'_3$. In order to simplify our analysis, we investigate the system as a function of $\epsilon'_3$, so that we can directly control the appearance of the dark state. 

 Importantly, the polaron transformation renders the coupling to the bath $H_{\rm s,b}^{(p) }$ weak in comparison to the  system parameters, so that the application of the common Redfield equation formalism is justified~\cite{Schaller2014a}. In doing so, we obtain  equations of motions for  the reduced density matrix of the system $\rho= \text{ Tr}_{\text{ br } }\left(\rho_{\text{tot}} \right)$, where $\rho_{\text{tot}}$ denotes the density matrix of the  system plus the reservoirs and the bath, and $\text{ Tr} _{\text{ br} }\left(. \right)$ denotes a trace over the bath and reservoirs' degrees of freedom. The equation of motion of the reduced density matrix finally reads
\begin{equation}
\frac{d}{dt} \rho = \mathcal W \rho,
\label{eq:masterequation}
\end{equation}
where $\mathcal W$ is a time-independent matrix and  denotes the Liouvillian superoperator. Details of the derivation can be found in Appendix~\ref{sec:masterEquation}. This approach can describe  both, incoherent and coherent processes~\cite{Xu2016}, similar to the method considered in Ref.~\cite{Gurvitz2016}. In contrast to Ref.~\cite{Entin-Wohlman}, the polaron transformation allows to study the influence of decoherence even for strong couplings $\gamma$.

\subsection{Transport observables}

\begin{figure}[t]
\includegraphics[width=\linewidth]{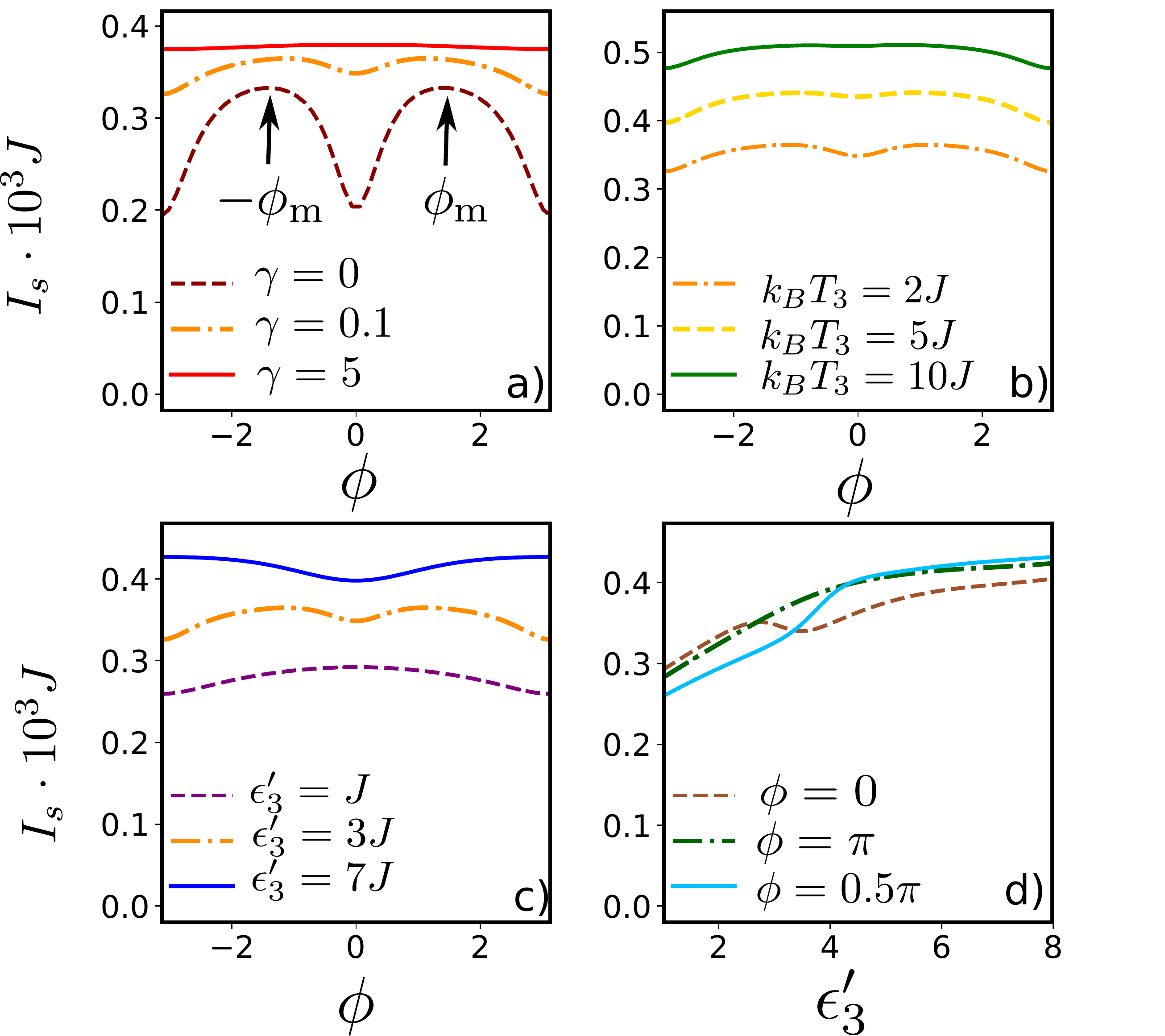}
\caption{ Steady-state currents $I_{s}$ entering reservoir $n=2$. If not specified, overall parameters are $J_1=J_2= J_3= J$, $\epsilon_1=\epsilon'_3= 3J$, $\epsilon_2=4J$, $\Gamma_1=\Gamma_2=0.05J$, $k_B T_n = 3 J$, $\mu_1=3.5J$, $\mu_2= 1.5J$,  $\gamma=0.1$, $\omega_{\text c}= 10 J$ , $\Gamma_{n}^{(0)}= 0.005 J$, $\omega_{\text c,n}= 10 J$ and  $\omega_{0 ,n}= 2 J$ for $n=1,2$.
}
\label{fig:currents}
\end{figure}%
%
\begin{figure}[t]
\includegraphics[width=\linewidth]{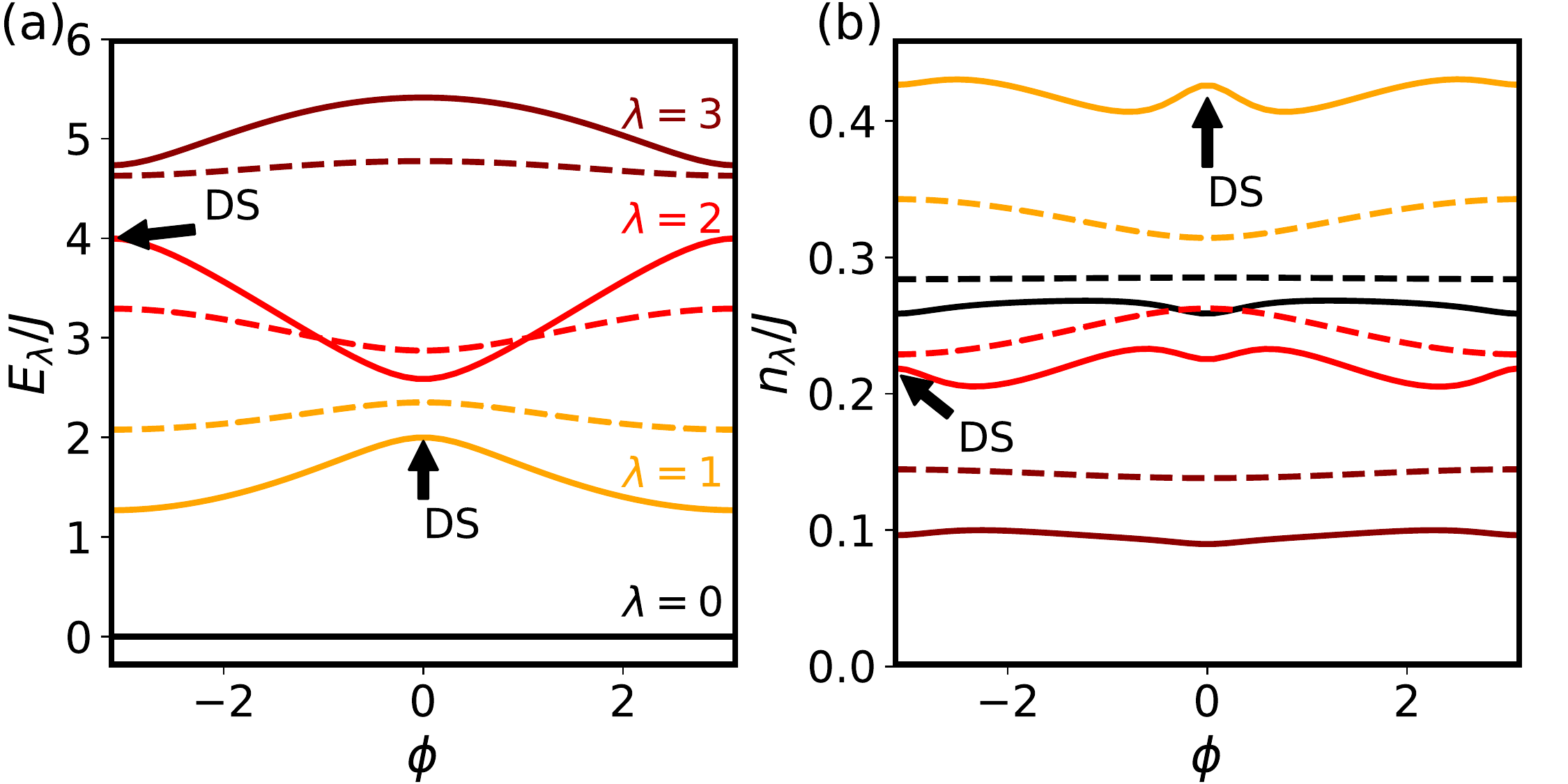}
\caption{ Spectral analyis of Fig.~\ref{fig:currents}(a) for $\gamma=0$ (solid lines) and $\gamma=5$ (dashed lines). In (a) we depict the spectrum of the polaron-transformed system Hamiltonian and in (b) we depict the occupation of the eigenstates $\lambda=0,1,2,3$ in the stationary steady state. $\lambda=0$ corresponds to the empty state. We mark the dark states for $\gamma=0$ (DS) with arrows. The dark states for $\gamma=5$ are located on corresponding positions.
}
\label{fig:spectralAnalysis}
\end{figure}%
%

Equation~\eqref{eq:masterequation}  can be amended to take account of transport statistics. In doing so, we consider the conditioned density matrix $\rho(t, \mathbf N)$ with $\mathbf N = (N_1,N_2)$ of the central system, which  contains additional information about the number of particles $N_n$ in  reservoir $n=1,2$. To this end, we unravel Eq.~\eqref{eq:masterequation}  as
\begin{equation}
	\frac{d}{dt}\rho(t, \mathbf N) = \mathcal W_{0} \rho(t, \mathbf N) + \sum_{n,\sigma=\pm} \mathcal W_{\sigma}^{n}  \rho (t, \mathbf N^{(n,-\sigma)}    ),
\end{equation}
where $\mathcal W_{0}$ describes a time evolution with no particle jumps from or to the reservoirs. The superoperators $\mathcal W_{\sigma}^{n}$ add $( \sigma=1)$ or remove $(\sigma=-1)$ a particle to the system, while correspondingly remove or add it to the reservoir $n$. The quantity $\mathbf N^{(n,\sigma)} $ is equal to $\mathbf N$ but with $N_{n}$ replaced by $N_{n}+\sigma$. 

Applying a Fourier transformation in the particle space $\mathbf N=(N_1,N_2)$, the generalized reduced density matrix $\rho(t, \boldsymbol \chi ) $ fulfills the equation of motion 
\begin{align}
\frac{d}{dt}\rho(t, \boldsymbol \chi )  &= \mathcal W (\boldsymbol \chi)\rho(t, \boldsymbol \chi ) \newline\nonumber \\
  &=  \mathcal W_{0} \rho(t, \boldsymbol \chi ) + \sum_{n,\sigma=\pm}       e^{i \sigma \chi_{n} }\mathcal W_{\sigma}^{n}  \rho (t, \boldsymbol \chi  ),
\end{align}
where $\boldsymbol \chi =(\chi_1,\chi_2)$ are the variables conjugated to  $\mathbf N$.
The current and the frequency-dependent noise can be calculated using
\begin{align}
 i I_{n} &= \left< \mathcal J_{n}^{(1)}  \right>,  \\
 i^2 S_{n}^{(2)} &=  \left< \mathcal J_{n}^{(2)}  + \mathcal J^{(1)}_{n}\left[  \Omega_{0} (i \omega) + \Omega_{0} (- i \omega) \right]\mathcal J_{n}^{(1) } \right>,
 \label{eq:noise}
\end{align}
where $\mathcal J_{n}^{(m)} \equiv \partial_{\chi_{n} }^{m} \mathcal W (\boldsymbol \chi) $ and $\Omega_{0} (z)\equiv \left[ z -\mathcal W_0 \right] ^{-1} $~\cite{Marcos2010,Liu2017}.

\begin{figure}[t]
\includegraphics[width=0.8\linewidth]{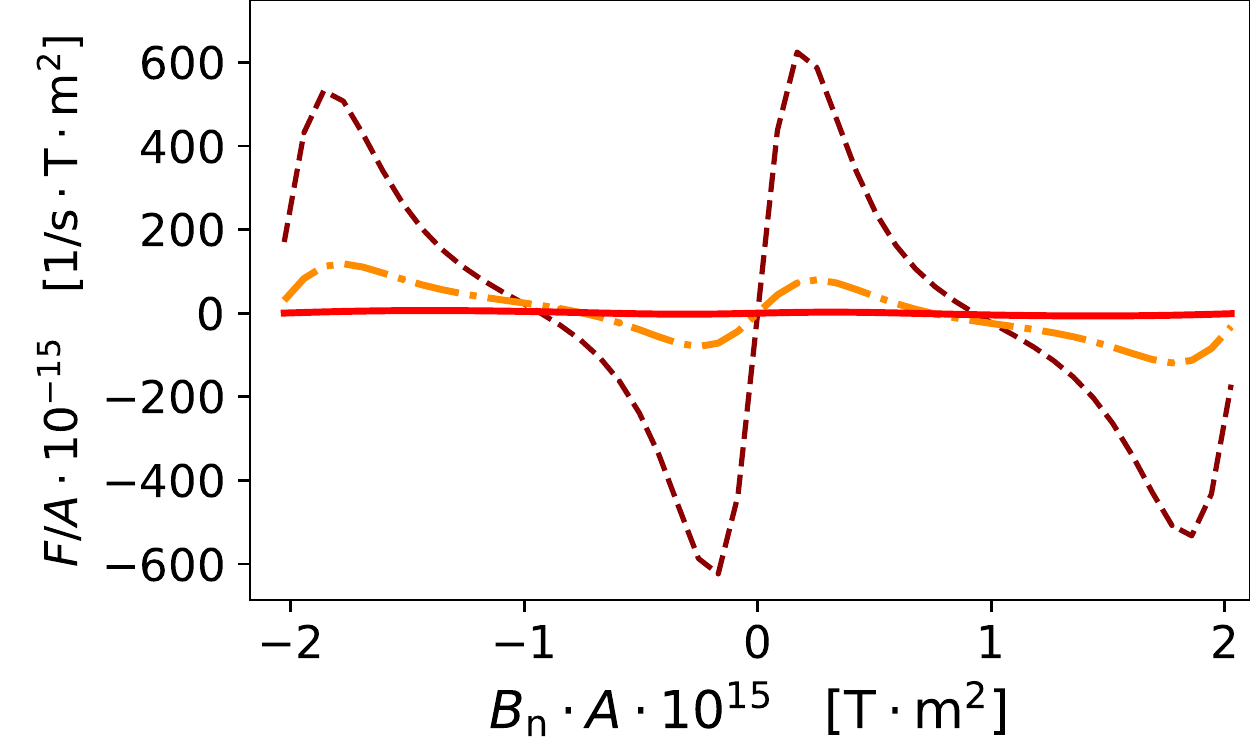}
\caption{ Figure of merit $F\equiv \partial I_{\rm s}  /\partial B_{\rm n} $ defined as the derivative of the stationary current with respect to the magnetic field component normal to the ring scaled by the area of the ring $A$. The curves correspond to the tow cases  in Fig.~\ref{fig:currents}(a). We have chosen an experimental realistic value of $\Gamma_{n}^{(0)}= 10\,\rm k Hz$.
}
\label{fig:figureOfMerit}
\end{figure}%
%

\begin{figure}[t]
\includegraphics[width=\linewidth]{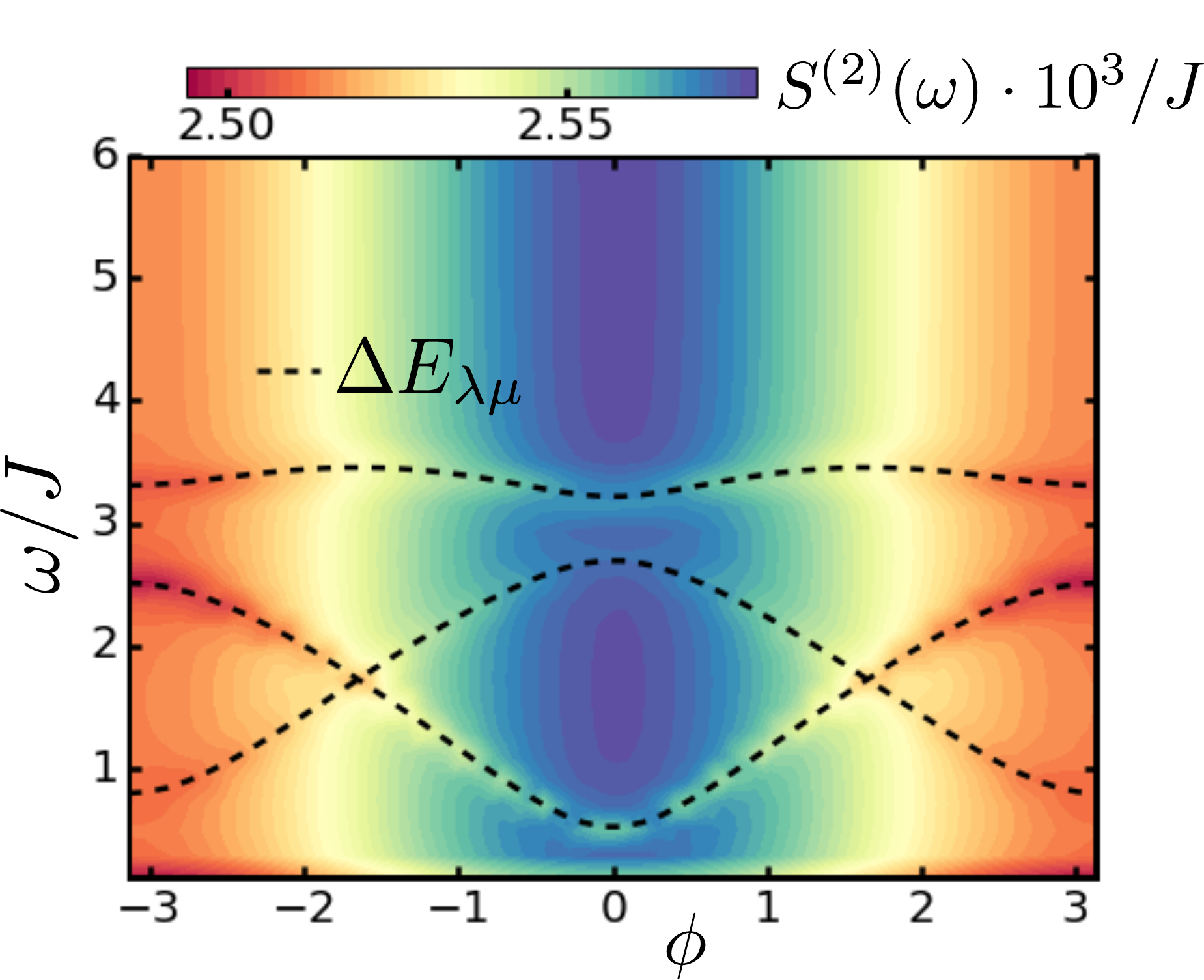}
\caption{ Frequency dependent noise. The parameters are as in Fig.~\ref{fig:currents}(a) for $\gamma=1$. Dashed lines depict the energy differences of the three eigenstates $\Delta E_{\lambda\mu}=E_{\lambda}-E_{\mu}$ of the polaron-transformed system, with $\lambda,\mu=1,2,3$.
}
\label{fig:noise}
\end{figure}%
%

\begin{figure*}[t]
\includegraphics[width=\linewidth]{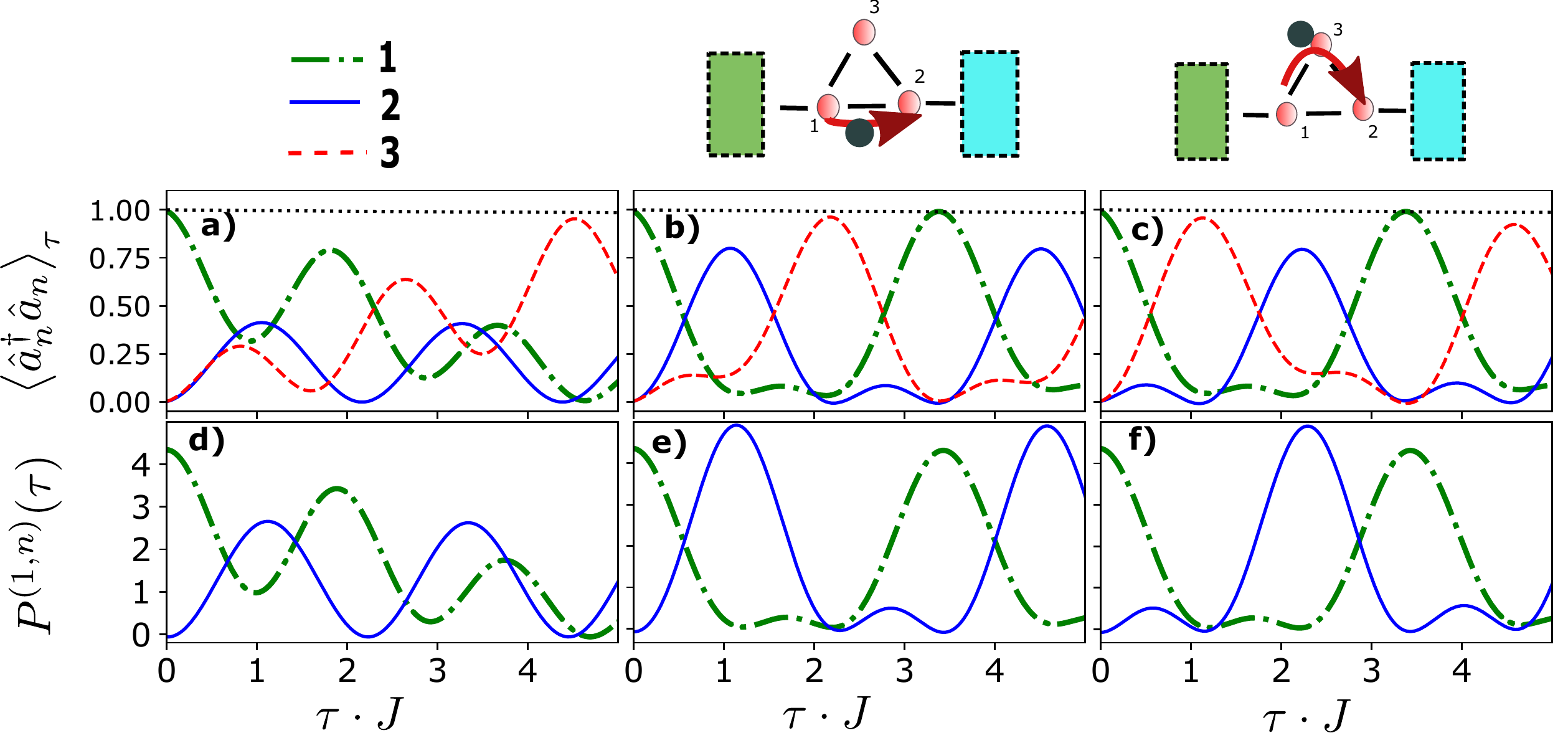}
\caption{ (a)-(c) Occupation of the sites $n=1,2,3$ (dotted-dashed, solid, dashed) as a function of time $\tau$ for a charge which enters the  system in Fig.~\ref{fig:overview} from reservoir $n=1$ for different $\phi$. The black dotted line depicts the total occupation of all sites. (d)-(f) Waiting time distribution $P^{(1,n)}( \tau)$ that   the particle jumps into reservoir $n=1$ or $n=2$ after the  time difference $\tau$, as defined in Eq.~\eqref{eq:waitingTimeDistribution}. Overall parameters are as in Fig.~\ref{fig:currents}(a), but $\gamma=0$, and $k_{\rm B}T_n=5J$.  The phases are $\phi=0$ ((a),(d)), $\phi=\pi$ ((b),(e)), and $\phi=-\pi$ ((c),(f)).
}
\label{fig:waitingTime}
\end{figure*}%
%

As we see later, the statistical distribution $P^{(n,n')}( \tau)$ of the  time difference $ \tau$ between two consecutive jump events $\mathcal W_{\sigma=1}^{n}$ and $\mathcal W_{\sigma'=-1}^{n'}$ contains interesting information about the system dynamics~\cite{Engelhardt2017}. 
This waiting-time experiment is sketched in Fig.~\ref{fig:overview}(b). At time $t=t_0$, a particle enters the system from reservoir $n$. The waiting time distribution describes the probability that the particle jumps into reservoir $n'$ at time $t_1=t_0+\tau$. In between, the dynamics is governed by the internal properties of the system. Thus, the waiting time distribution can provide insight to the AB effect and the dephasing dynamics.   

According to Ref.~\cite{Brandes2008}, the waiting-time probability distribution can be calculated using
\begin{equation}
	P^{(n,n')}( \tau ) = \frac{\text{Tr}\left(\mathcal W_{-1}^{n'}  e^{ \tau \mathcal W_{0} } \mathcal W_{1}^{n}\rho_s  \right)    }{\text{Tr}\left(\mathcal W_{1}^{n}\rho_s  \right)},
	\label{eq:waitingTimeDistribution}
\end{equation}
where $\rho_s$ denotes the steady states of the system with $\dot \rho_{s}=0$ for the dynamics in Eq.~\eqref{eq:masterequation}. The denominator ensures the normalization
\begin{equation}
\sum_{n'} \int_{0}^{\infty} P^{(n,n')}( \tau)d\tau = 1.
\end{equation}
Single-electron resolution in transport has been achieved in single-electron transistor experiments such as in Refs.~\cite{Flindt2009,Wagner2017}. Moreover,  waiting-time distributions have been suggested to reveal the interplay of the electronic and vibrational degrees of freedom in  molecular junctions~\cite{Kosov2017}. 

\section{Results}

\subsection{Steady-state current}

In Fig.~\ref{fig:currents}, we plot the stationary current as a function of the phase $\phi$ for different system parameters. In Fig.~\ref{fig:currents}(a), we focus on the influence of the system-bath coupling,  characterized by the spectral coupling density
\begin{equation}
	\Gamma_{b}(\omega) \equiv \sum_{k} \left|  V_k \right|^2 \delta(\omega- \omega_{k}) = \gamma \frac{\omega^3}{\omega_c^2} e^{- \omega/ \omega_c},
\end{equation} 
where we have chosen  a super-Ohmic parametrization with cut-off frequency $\omega_c$. 

For the coupling strength $\gamma=0$ we observe that the current $I_{s}$ sensitively depends on the phase $\phi $.  It is interesting to see that the current exhibits a local minimum at $\phi=0$, while the maximum current is reached here for finite phases $\phi=\pm \phi_{\text m}$ for the chosen chemical potential of $\mu_{1,2}$.  In other words, transport can be enhanced with the help of the AB effect, i.e., $\phi\neq 0$.

As the system-bath coupling strength $\gamma$ increases,  the $\phi$ dependence of the current becomes weaker. A rough explanation of this behavior is that the AB effect is based on the coherent wave nature of the charge, i.e., the interference of the two pathways. Due to the increasing coupling to the thermal bath, thermal fluctuations increasingly destroy the coherent dynamics. This behavior can be observed in Fig.~\ref{fig:currents}(b), where a raising temperature leads to enhanced thermal fluctuations, which suppress the coherent system dynamics and consequently the $\phi$ dependence.

The phase dependence of the current can be understood as follows. According to the Redfield equation with the Born-Markov secular approximation, the expression for the current through the system can be written as
\begin{align}
	I_{s}=& \sum_{\lambda=1,2,3} I_{s}^{(\lambda)},\\%
	I_{s}^{(\lambda)} &= P_{0} \frac{k_{1,\lambda}^{+}  k_{2,\lambda}^{-} -   k_{1,\lambda}^{-}  k_{2,\lambda}^{+}  }{   k_{1,\lambda}^{-}   +  k_{2,\lambda}^{-}   },
	\label{eq:CurrentLambda0}
\end{align}
where
\begin{align}
k_{n,\lambda}^{+}= \left|\left< \lambda \right| \hat a ^{\dagger}_{n}  \left|\text{vac} \right> \right|^{2} f_n(E_{\lambda}) \Gamma_n(E_{\lambda})= k_{n,\lambda}^{-} e^{\frac { E_{\lambda}-\mu_{n}}{k_{\text B} T_{n}}}
\label{eq:transitionRate}
\end{align}
denotes the transition rates from the vacuum state to the eigenstates $\left| \lambda\right>$ with energy $E_{\lambda}$, induced by  reservoir $n$. $P_0$ denotes the probability that the system is in the vacuum state. Here, $f_n(E)$ is the Fermi distribution in reservoir $n$ and $\Gamma_n(\omega)=\Gamma_{n}^{(0)}\omega_{\text c,n}^2  / \left[ (\omega-\omega_{0,n} )^2+\omega_{\text c,n}^2\right] $.

In  appendix \ref{app:CurrentFormula} we show that the current through the eigenstate $\left| \lambda \right>$ in Eq.~\eqref{eq:CurrentLambda0} can be written as
\begin{equation}
I_{s}^{(\lambda)} =  \mathcal N_\lambda \left|J_{1} \cdot  (E_{\lambda}- \epsilon'_3 ) +  J_{2} J_{3} e^{ \phi} \right|^2.
\label{eq:apprCurrent1}
\end{equation}
This expression illustrates the $\phi$ sensitivity and its dependence on the thermal fluctuations. For an increasing system-bath coupling $\gamma$ or for a large temperature $T_3$, the tunneling parameters are renormalized as $J_{2}\rightarrow \kappa J_{2}$ and $J_{3}\rightarrow \kappa J_{3}$ according to Eq.~\eqref{eq:renormalization}. For larger $\gamma $ or $T_3$ the polaron renormalization parameter $\kappa$ approaches $0$. In that limit, it is clear from Eq.~\eqref{eq:renormalization} that the coherence in the polaron-transformed Hamiltonian is suppressed so that the AB effect vanishes.
Accordingly, in Eq.\eqref{eq:apprCurrent1} we find  that the $ I_{s}^{(\lambda)}$ dependence on $\phi$ vanishes. A similar reasoning applies to the factor $\mathcal N_\lambda$.
The thermal bath can thus be used to tune the AB effect and the related time-reversal symmetry breaking. This is thus reminiscent to the breaking of spatial symmetries with dephasing as investigated in Ref.~\cite{Thingna2016}.

From  Eq.~\eqref{eq:apprCurrent1} it is hard to determine the optimal phase $\phi_{\text m}$ for the maximum current as  the $\phi$ dependence of $\mathcal N_\lambda$ is complicated, as shown in appendix~\ref{app:apprFormula}.
Moreover, Eq.~\eqref{eq:apprCurrent1} reveals that the current strongly depends on the onsite potential $\epsilon'_3$. In Fig.~\ref{fig:currents}(c) we depict the phase dependence of the current for different values of $\epsilon'_3$.  The $\phi$ dependence is most significant if $\epsilon'_3$ is on the same order as $\epsilon_1$ and $\epsilon_2$. 
If $\epsilon'_3$ is detuned from the other onsite-energies, the site $n=3$ is hard to reach for a charge initially located at one of the other sites. For this reason, the intra-ring coherence and therefor the AB effect is weakened. Also the $\phi$ dependence looks different depending on $\epsilon'_3$.  For a large (small) $\epsilon'_3$, we observe a minimum (maximum) current  $\phi=0$, and a maximum (minimum) current at $\phi=\pi$. For $\epsilon'_3\approx \epsilon_1,\epsilon_2$,  the maximum current is found at finite $\phi=\pm \phi_{\rm m}$ and exhibits a local minimum at $\phi=0$. Thus, by tuning $\epsilon'_3$, one can control the current-phase dependence.

Furthermore, in Fig.~\ref{fig:currents}(d) we find that by adjusting the phase  $\phi$, we can generate distinct $\epsilon'_3$ dependencies for the current. For all $\phi$ we observe that the current for small $\epsilon'_3$ is lower than for large $\epsilon'_3$.  This happens as we allow here only  for single-electron occupation of the system: If $(\epsilon'_3\ll\epsilon_1,\epsilon_2) $ or $(\epsilon'_3\gg \epsilon_1,\epsilon_2 )$ then there is an eigenstate strongly localized at $n=3$. For $(\epsilon'_3\ll\epsilon_1,\epsilon_2 ) $ this state is likely occupied with a charge, so that the transport is blocked. For $(\epsilon'_3\gg \epsilon_1,\epsilon_2) $, the localized state is likely to be empty, so that the transport properties are determined by the coherent transition from sites $1 \rightarrow 2$. 
In particular, for $\phi=0$ we observe a non-monotonic dependence on  $\epsilon'_3$. This can be explained with the appearance of a dark state as explained in the following subsection.
\begin{figure*}[t]
	\includegraphics[width=\linewidth]{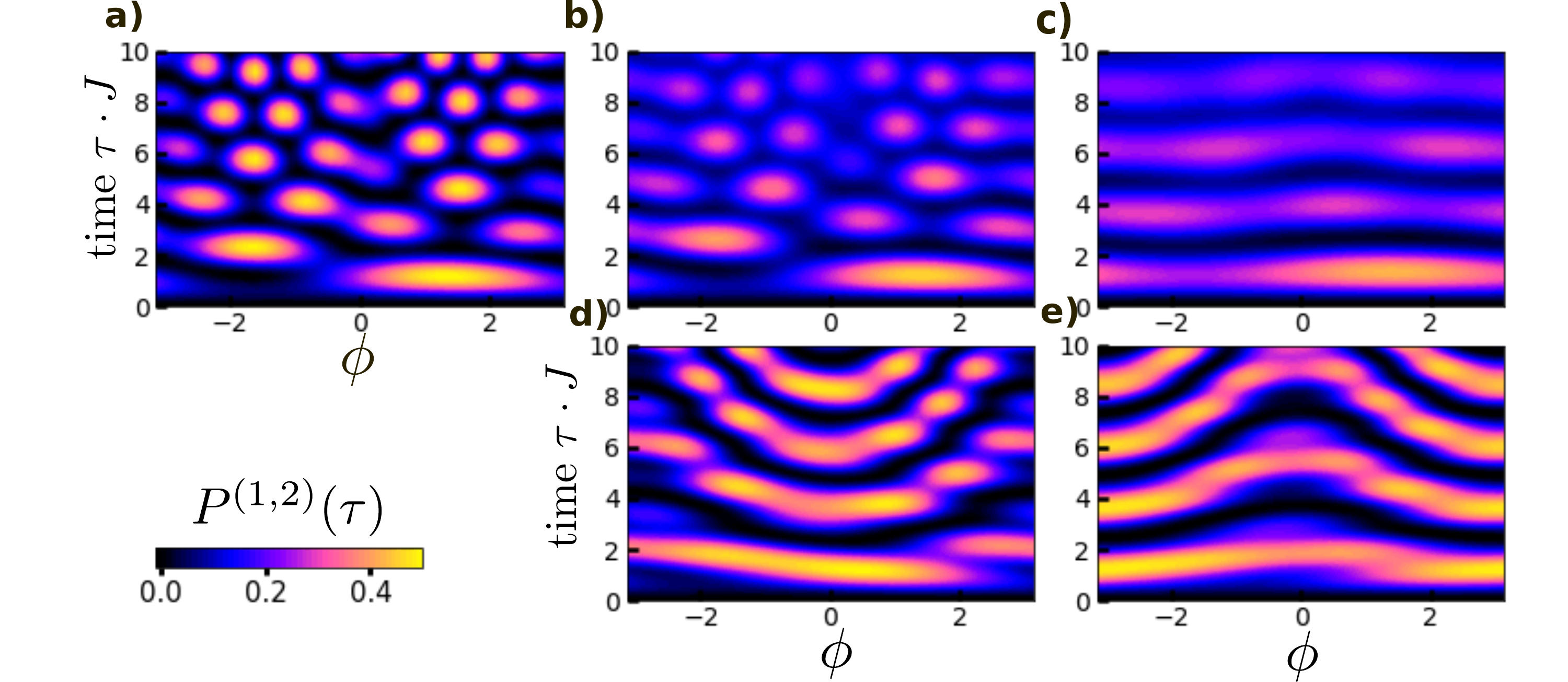}
	\caption{ Waiting time distribution $P^{(1,2)}( \tau)$ for a particle which enters the central system in Fig.~\ref{fig:overview}(a) from reservoir $n=1$  and jumps to reservoir $n=2$  as defined in Eq.~\eqref{eq:waitingTimeDistribution}. Overall parameters are the same as in Fig.~\ref{fig:currents}(a), but $k_{B}T_n=5J$. In panels (a), (b), and (c), we have chosen $\gamma=0$, $\gamma=1$, and $\gamma=5$, respectively, and $\epsilon'_3=3J$. In panels (d) and (e), the parameters are $\epsilon'_3=J$ and $\epsilon'_3=7J$ for $\gamma=0$.  
	}
	\label{fig:waitingTimePD}
\end{figure*}%
\subsection{Role of dark state}

The current characteristics can be explained following the arguments in Ref.~\cite{Cao2009}. It is strongly related to the appearance of the dark state as described in Sec.~\ref{sec:darkState}. Inserting the energy for the dark state $E^{\rm dark}_{0,\pi}$ below Eq.~\eqref{eq:darkStateWF} into the partial current expression Eq.~\eqref{eq:apprCurrent1} for  $\phi =0,\pi$, we find that $I_{s}^{(\lambda)}=0$, i.e., the dark state does not contribute to the current. Upon tuning the phase away from $\phi=0$ or $\pi$, the dark state disappears, so that all eigenstates contribute to the current.
Thus, the dark state may give rise to a  suppression of the transport. 

To understand this in  detail, we depict in Fig.~\ref{fig:spectralAnalysis}(a) the spectrum $E_\lambda$ of the polaron-transformed system Hamiltonian, and in Fig.~\ref{fig:spectralAnalysis}(b) the occupation of the corresponding eigenstates $\lambda$. The dark states  appear for   $\phi=0$ and $\phi=\pi/2$ in eigenstate $\lambda=1$ and $\lambda=2 $, respectively. Both are energetically located on a local maximum of $E_{\lambda}(\phi)$. For this reason, one might conclude, that they are less thermally occupied in comparison to the other $\phi$ values. Yet, as we observe in Fig.~\ref{fig:spectralAnalysis}(b), the occupation $n_\lambda$ of the eigenstate $\lambda$ exhibits a local maximum, as the dark state Eq.~\eqref{eq:darkStateWF} is only coupled to reservoir $n=1$, but not to reservoirs $n=2$. Thus, a charge occupying the dark state can neither enter reservoir $n=1$, because  all orbitals are almost occupied due to the chosen chemical potential $\mu_1$,  nor can it enter reservoir $n=2$, as the dark state has no overlap with site $n=2$. This leads to an enhanced occupation of the dark state and to a blockade of the transport.

The presence of a dark state  explains  the dependence of the current on $\phi$ and $\epsilon'_3$,  in Fig.~\ref{fig:currents}(a) and (d). As the dark state only appears at singular parameter combinations, e.g., $\phi=0$ and $\epsilon'_3=\epsilon_1$, the current shows a  decrease when the system parameters approach the dark state configuration. 
In Fig.~\ref{fig:currents}(d), the influence of the dark state is not as prominent for $\phi=\pi$ as for $\phi=0$, because  the dark state energy $E^{\text{dark }}_{\pi} =  \epsilon_1 + J$  is above the transport window $E^{\text{dark }}_{\pi}> E \in (\mu_{2},\mu_{1})$. Nevertheless the current for $\phi=\pi$ is significantly smaller than the current for $\phi=\pi/2$, which is a consequence of the dark state.

We emphasize that the dark state blockade and the minimum in the transmission  observed in the molecular transport experiment~\cite{Guedon2012} are both based on destructive interference of different charge paths. The experiment demonstrates that the electronic transport occurs essentially via three localized molecular orbitals, so that it is strongly related to our model system. Furthermore, interference experiments related to dark states have interesting applications such as the interaction free measurements using anti-resonances in optical and solid-state systems, which are also feasible to coherently detect dissipation~\cite{Kwiat1995,Chirolli2010,Strambini2010}.

\subsection{Figure of merit}

In an experimental investigation, the response of the current due to a change of the magnetic field is important. To this end, we define a figure of merit by
\begin{equation}
 F = \frac{\partial I_s}{\partial B_{\rm n}},
\end{equation} 
which we depict, scaled by the ring area $A$, in Fig.~\ref{fig:figureOfMerit}. Here we use the same parameters as in Fig.~\ref{fig:currents}(a), and we choose an experimentally realistic reservoir coupling of $\Gamma_{n}^{(0)}=10\, \rm kHz$~\cite{Flindt2009}.  From Fig.~\ref{fig:currents}(a) we can thus conclude that $I_{\rm s}(\phi\approx0.5) \approx 500 \frac 1{\rm s} $. In the following we discuss two examples.

A semiconductor quantum dot (or single electron transistor) has a characteristic size of $1 \rm \mu m$, so that  the area of a triple quantum dot is in the order of $A=1 \rm \mu m^2$. $\phi=0.5$  corresponds  to a magnetic field of $B_{\rm n} \approx 0.3 \,\rm mT$. From Fig.~\ref{fig:figureOfMerit}, we find that for this magnetic field the figure of merit is  $F= 6 \times 10^{5} \frac{1}{\rm s T}$, so that an increase  of $\Delta B_n = 0.01\, \rm mT$ gives rise to a current change of $\Delta I_s \approx 6 \frac 1{\rm s}$. This can be measured in experiments \cite{Flindt2009}. 

Molecules can have ring structures with area  $A\approx1\, \rm n m^2$, so that a magnetic field of $B_{\rm n}= 300\,\rm T$ is needed to acquire a phase of $\phi=0.5$. However, the largest artificially generated continuous magnetic field is nowadays $B_n= 45\,\rm T$~\cite{magneticField}. Thus, the values $\phi\approx 0,\pi$ in our model refer to  interference phenomena in molecular ring structures, such as experimentally investigated in Ref.~\cite{Guedon2012,Vazquez2012}. Our investigation based on the polaron transformation helps to reveal the impact of dephasing on  the interference in molecular setups, where the coupling between electronic and vibronic degrees of freedom is crucial.

\subsection{Frequency dependent noise}

Although the current shows a strong dependence on the phase $\phi$, it is symmetric regarding the inversion $\phi\rightarrow -\phi$. This symmetry is denoted as phase rigidity~\cite{Tu2012}. It is related to the fact that the current is not sensitive to the details of the coherent dynamics of the system. To get more information about the internal dynamics, the frequency dependent noise Eq.~\eqref{eq:noise} is a more appropriate observable. We depict the noise in Fig.~\ref{fig:noise} as a function of $\phi$ and $\omega$. We observe a mainly frequency independent pattern which results from the $\mathcal J^{(2)}_{n=2}$ operator in Eq.~\eqref{eq:noise}. On top of that, we observe a frequency-dependent structure. This structure resembles the energy differences of the eigenstates $\Delta E_{\lambda\mu}=E_{\lambda}-E_{\mu}$, which we depict in Fig.~\ref{fig:noise} with  dashed lines. Although it thus gives  information about the internal structure of the system, it nevertheless  exhibits a $\phi\leftrightarrow -\phi$ symmetry.

We stress that the phase rigidity is a consequence of the fact that our system exhibits two electronic reservoirs. It is well-know that the phase rigidity in the stationary current is broken when increasing the number of electronic reservoirs, as experimentally demonstrated in Ref.~\cite{Kobayashi2002,Strambini2009}. In contrast, the third terminal in our system is a heat bath, which does not break the phase rigidity, even not for strong coupling, as demonstrated here.

\subsection{Waiting time statistics}

In contrast to the stationary current and its noise, the waiting time statistics as defined in Eq.~\eqref{eq:waitingTimeDistribution} indeed violates the $\phi\leftrightarrow -\phi$ symmetry, similar to the transient current investigated in Ref.~\cite{Tu2012}.  Figure~\ref{fig:waitingTime} shows instances of time evolutions  for  $\phi=0,\pm \pi/2 $. In Fig.~\ref{fig:waitingTime}(a)-(c) we plot the occupations of the sites $N_{n}$ for $n=1,2,3$ by a particle which has jumped at time $t=t_0$ from reservoir $n=1$ into the system.

 In all three cases we observe an oscillating time evolution of the site occupations. The total system occupation, which is initially $\sum_nN_{n}=1$, decreases as a function of time. This reflects  that the charge can leave the three-site system and enters one of the reservoirs. Interestingly, for $\phi=+\pi/2$, the particle  first visits site $n=2$ and then  $n=3$. In contrast, for $\phi=-\pi/2$, the particle  first occupies site $n=3$ before it goes to $n=2$. Thus,   the phase $\phi$ steers the path of the particle.
 
In Fig.~\ref{fig:waitingTime}(d)-(f), we plot  the corresponding waiting-time distributions that the particle jumps after time  $\tau$  into reservoir $n=1,2$. The waiting-time distribution function are strongly correlated to the occupations of the sites $n=1,2$. Thus, the waiting time experiment allows to study the internal population dynamics of the system.

In Fig.~\ref{fig:waitingTimePD} we investigate the waiting-time distribution $P^{(1,2)}( t)$ as a function of $\phi$ for different parameters. 
 In Fig.~\ref{fig:waitingTimePD}(a) we consider the case $\gamma=0$, which exhibits an interesting pattern that breaks the $\phi\leftrightarrow -\phi$ symmetry. In particular, for $\phi>0$ the waiting time distribution reaches faster a maximum than for a negative $\phi$. This is in agreement with the explanations of Fig.~\ref{fig:waitingTime}.

Figures~\ref{fig:waitingTimePD}(b) and (c) depict the influence of dephasing on the waiting time statistics. In Fig.~\ref{fig:waitingTimePD}(b), we consider an intermediate dephasing parameter $\gamma=1$ and observe that the pattern is essentially equivalent to Fig.~\ref{fig:waitingTimePD}(a), but  more smeared by decoherence. 

The observations in Fig.~\ref{fig:waitingTimePD}(c), where we consider a strong system-bath coupling $\gamma = 5  $,   is remarkable. Although the $\phi$ dependence has almost vanished for large times $\tau \cdot J >4$,  the $\phi$ dependence  is still clearly pronounced for short times. 
Here, the influence of the heat bath $n=3$ onto the dynamics becomes particular obvious. A positive $\phi$  directs the particle from site $n=1$ to $n=2$ at short times, so that the short-time waiting-time distribution is  reminiscence to the $\gamma=0$ case.  A negative $\phi$ directs the particle to the dephasing site $n=3$, so that the action of the dephasing is  visible even at short times.

In Fig.~\ref{fig:waitingTimePD}(d) and (e) we investigate the influence of the onsite potential $\epsilon'_3$ by considering $\epsilon'_3=J$ and $\epsilon'_3=7J$ for $\gamma=0$. Although the waiting time distribution is still sensitive on the phase, the broken  symmetry $\phi\leftrightarrow -\phi$ is less obvious then in Fig.~\ref{fig:waitingTimePD}(a). This is a consequence of the large detuning of $\epsilon'_3$ in comparison to $\epsilon_1=3J$ and $\epsilon_2=4J$, which makes the transfer over $n=3$ unlikely. Thus, the ring structure, an essential ingredient for the AB effect, is weakened. As a consequence, the isolated probability maxima observed in Fig.~\ref{fig:waitingTimePD}(a) become horizontally extended and connected, so that they form a band-like pattern as we can see clearly in  Fig.~\ref{fig:waitingTimePD}(d). This resembles the decreased dependence on $\phi$.
A similar pattern can be  found in \ref{fig:waitingTimePD}(e), at which the bands are more deformed.

\subsection{Tuning the interference with the Aharanov-Bohm phase}

\label{sec:pathInterference}

To analytically evaluate  the interference effect, we distribute the dynamics of the charge into the two path ways which the particle can take as shown in the sketches of Fig.~\ref{fig:waitingTime}. The probability that the charge is located on site $n=2$  at time $t$ with the initial population at  $n=1$ at time $t=0$ is given by
\begin{equation}
	\left| \left< 2\right| \hat U (t) \left|1 \right>\right|^2 \approx
	\left| \left< 2\right| \hat  U^{(0)} (t) + \hat U^{(1)} (t) \left|1 \right>\right|^2,
	\label{eq:pathInterference}
\end{equation}
where  the time evolution operator $\hat U (t)$ of the system is approximated  by a sum of two time evolution operators  $\hat U^{(0)}$ and $\hat U^{(1)}$. These two operators are related to $ \hat U (t)$ by replacing either $J_{3}=J_{2}\rightarrow 0$  or $J_{1}\rightarrow 0 $, respectively. 

For simplicity, we assume here that $\epsilon_n=\epsilon$, and $J_{n}=J$. The calculation of the right hand site of Eq.~\eqref{eq:pathInterference} can be performed analytically as shown in Appendix \ref{app:pathInterference}. We find
\begin{equation}
	\left| \left< 2\right| \hat U (t) \left|1 \right>\right|^2 \approx
\left|	i e^{i \phi } \sin(J t)-\frac 12 + \frac 12 \cos(\sqrt{2}J t ) \right|^2.
\end{equation}
This expression allows for an intuitive interpretation of the interference pattern observed in Fig.~\ref{fig:waitingTimePD}(a)-(c). We find a maximum constructive interference for $\phi = \frac \pi 2$, while for $\phi= -\frac{ \pi}{2}$ we observe a destructive interference for short times. Moreover,  for $\phi = \frac \pi 2$ the amplitude is maximal at $J \cdot t\approx \frac \pi 2 $, while  for $\phi = - \frac \pi 2$ it is maximal for $J \cdot t\approx  3\frac \pi 2 $. Although we have assumed that $\epsilon_n$ are all equal, this is in good agreement with the observation in Fig.~\ref{fig:waitingTimePD}(a)-(c).

However, this approach fails to describe the pattern in Fig.~\ref{fig:waitingTimePD}(d)-(e). In the approximation Eq.~\eqref{eq:pathInterference}, we have assumed that both paths are equally likely. If $\epsilon'_3$ is detuned from  $\epsilon_1$ and $\epsilon_2$, then the charge will prefer the direct way $1\rightarrow 2$, as this constitutes a resonant particle transfer. This preference is not incorporated in Eq.~\eqref{eq:pathInterference}.

\begin{figure}[t]
	\includegraphics[width=\linewidth]{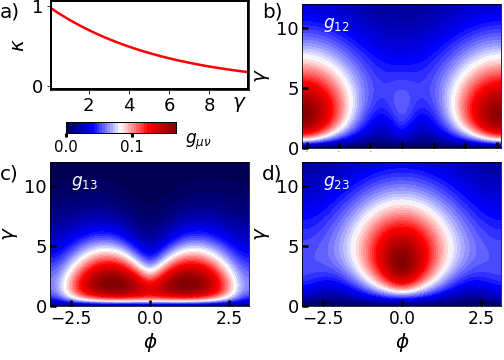}
	\caption{ (a) Renormalization factor $\kappa$ as a function of $\gamma$. (b) Transition rates $\gamma_{\mu,\nu}= \mathcal W_{\mu\mu,\nu,\nu}$ between the eigenstates $\mu,\nu=1,2,3$ induced by the heat bath $n=3$. Parameters are equal to Fig.~\ref{fig:waitingTimePD}(a).
	}
	\label{fig:dissipation}
\end{figure}%

\subsection{Aharonov-Bohm effect vs. dephasing}

Figures~\ref{fig:waitingTimePD} (a)-(c) shed new light on the dynamics of the three-site model. Let us compare two observations.
First, right after the charge is localized at $t=t_0$, its dynamics is   sensitive to $\phi$ and evolves according to the system Hamiltonian and thus the AB effect, as can be seen in all panels in Fig.~\ref{fig:waitingTimePD}.
Second, the dephasing  does not occur immediately, but the phase information becomes gradually deleted and vanishes at long times $\tau \cdot J \approx 4$ in Fig.~\ref{fig:waitingTimePD} (c). Thus, in contrast to the rapid  onset of the  AB effect, the dephasing is a cumulative  effect. In particular, observing  Fig.~\ref{fig:waitingTimePD}(c), we  conclude that the dephasing takes place only when the particle passes site $n=3$. Our findings are consistent with Ref.~\cite{Chen2014}, which demonstrated that in transport through molecules with localized dephasing probe decoherence does not take place immediately.

In Fig.~\ref{fig:dissipation}, we investigate the impact of dephasing onto the system in more detail. In Fig.~\ref{fig:dissipation}(a) we depict the dependence of $\kappa$ on the  system-bath coupling $\gamma$. The exponential decrease according to Eq.~\eqref{eq:kappa} is determined by the temperature $T_3$. For $\gamma=5$ as in Fig.~\ref{fig:waitingTimePD}(c), we find a reduction of $50 \%$, so that the ring structure is still quite strong. In Fig.~\ref{fig:dissipation}(b)-(d), we investigate the transition rate $g_{\mu\nu}$ determining the thermally induced transitions between eigenstates $\mu$ to $\nu$ in the Redfield equation~\eqref{eq:masterequation}. For the rate $g_{12}$ at $\phi=\pi$ we find a crossover from an increasing rates for small $\gamma$ to a decreasing rate at large $\gamma$. The maximum can be found around $\gamma = 3  $. This is a typical feature for the crossover from week to strong coupling~\cite{Wang2014,Wang2015}. It is surprising to see in Fig.~\ref{fig:dissipation}(b)  that $g_{12}$ strongly depends on $\phi$ for $\gamma = 3 $. While it is maximal at $\phi=\pi$, it is minimal for $\phi=0$. In contrast, $g_{23}$ exhibits a complementary behavior.

The rates can be written as a product $g_{\mu\nu}=C(E_{\mu} -E_{\nu} ) \cdot A_{\mu\nu}$, where $C(x)$ denotes the bath correlation function which depends on the energy differences and $A_{\mu\nu}$ depends on the coupling operators Eq.~$\eqref{eq:couplingOperator}$ and the eigenstate wave function of the polaron-transformed system Hamiltonian.
 Considering the spectra for small and large $\gamma$ in Fig.~\ref{fig:spectralAnalysis}(a), we find that the spectral dependence on $\phi$ reduces for larger $\gamma$, as $\kappa$ decreases. Thus, we can infer that the strong $\phi$ dependence observed in Fig.~\ref{fig:dissipation} is caused by the wave function deformation.
 The rate of $g_{13}$ is relatively small compared to the other rates as the thermal activation energy $E_{3}-E_{1}$ is relatively large.

\section{Concluding discussion}

The reported study of the interplay of the celebrated AB effect and dephasing has revealed several intriguing findings:

(1) The interplay of the coherent AB effect subjected to dephasing is visible in the current through the ring system. As expected, the signatures of the AB effect in the steady-state current vanish in the strong dephasing limit, so that the amplitude of the AB oscillations, i.e. the phase dependence, is significantly reduced. This reduction can be found for both, an increasing system-bath coupling and an increasing bath temperature. Therefore, large thermal fluctuations randomize the coherent phase relation between different pathways and thus suppress the phase-dependence. Yet, surprisingly, the steady-state current can increase with the dephasing rate, which can be attributed to the dark-state effect summarized next.

(2) The AB phase can be used to control the interference of the two transfer pathways. For a moderate dephasing rate, the phase can be harnessed to increase the steady-state current such that the optimal phase for the maximum current appears at a finite value,  $\phi_{\rm m}\neq 0 $, for the chosen chemical potential. The interplay of constructive and destructive interference is particularly strong if the system exhibits a dark state for singular values of the phase.  Precisely because of the dark state effect, the phase dependence of the current changes its functional form as the site energy $\epsilon_3$ increases. As the dark state is an effect, which relies on the coherent nature of the charges, the impact of the dephasing is clearly manifested in the response of the dark state on the increasing system-bath coupling. The phase dependence of the current was significantly reduced, when tuning the system parameters away from the dark state configuration. This effect is particularly strong, if one chooses the chemical potentials of the leads so that the dark state is located in the transport window.

(3) Furthermore, unique signatures of the AB effect, including a breaking of the phase rigidity, can be observed in the waiting-time statistics, which is related to the transient dynamics of the system.  While the steady-state current exhibits the symmetry with respect to the sign of the phase, the phase rigidity breaks down in the waiting-time statistics such that the positive and negative phases lead to different transient currents, $I(\phi,t)\neq I(-\phi,t) $ for small $t$.  Interestingly, this effect survives in the waiting time statistics at short times even in the presence of a strong system-bath coupling. This observation reveals that the AB effect dictates the initial dynamics while dephasing gradually sets in later.  Besides dephasing, the AB impact can be weakened by detuning the on-site potential $\epsilon_3$. This leads to a significant structural change of the interference pattern in the waiting-time statistics. In contrast, the interplay of dephasing and AB effect results in a gradually \textit{smearing} of the original pattern. Overall one can thus conclude that the waiting-time statistics is more sensitive to the interplay of AB effect and dephasing than the stationary current.

\section{Acknowledgments}
The authors gratefully acknowledge financial support
from the China Postdoc Science Foundation (Grant No.: 2018M640054) Natural Science Foundation of China
(under Grant No.:U1530401), the NSF (Grant No.: CHE 1800301 and CHE 1836913/home/georg/Nutstore/Projects/currentProjects/trBreakPolaron/redproofs/changes.tex
/home/georg/Nutstore/Projects/currentProjects/trBreakPolaron/redproofs/requestedChanges.pdf) as well as inspiring discussion with Shmuel Gurvitz, Chen Wang, Jianhua Jiang, and Dazhi Xu.

\bibliography{mybibliography}
\appendix

\section{Polaron transformation}
\label{sec:polaronTransformation}

\subsection{General Hamiltonian}

A generalization of the Hamiltonian in Eq.~\eqref{eq:hamiltonian}  reads
\begin{eqnarray}
	H_{\rm tot}= H_{\rm s} + H_{\rm b } + H_{\rm sb}+ H_{\rm r } + H_{\rm sr},
	\label{eq:generalHamiltonian}
\end{eqnarray}
where
\begin{eqnarray}
H_{\rm s} &=& \sum_{n} \epsilon_n \hat a_n^{\dagger} \hat a_n + \sum_{n\neq m}  J_{mn} \hat a_m^{\dagger} \hat a_n \\\
  H_{\rm r }&=& \sum_{n k} \Omega_{nk} \hat c_{nk}^\dagger \hat c_{nk}  \\
   H_{\rm s, r }&=& \sum_{n k} \gamma_{nk} \hat c_{nk}^\dagger \hat a_{nk} + h.c.,
\end{eqnarray}
where $\hat a_{n} ^\dagger$ and $\hat c_{n,k} ^\dagger$ are fermionic creation  operators while $\hat b_{nk}^\dagger$ account for the phononic/photonic degrees of freedoms in the heat baths.

\subsection{Polaron transformation}

Here and in the following we assume that the system is only weakly coupled to the reservoir, but we want to allow for an arbitrary coupling strength to the heat baths. To get a valid analytical description which applies over arbitrary system-bath coupling strength, we perform a poloran transformation in the following. Thereby we refer to Ref.~\cite{Lee2015}. The polaron transformation is performed by applying the unitary operator
\begin{align}
	e^{\hat S} &= e^{\sum_{n,k}  \frac{g_{nk}}{ \omega_{nk} } \hat a_n^{\dagger} \hat a_n  \left(\hat b_{nk}^\dagger-\hat b_{nk} \right)  }\nonumber \\
	&=    \prod_{n}e^{\sum_{k}  \frac{g_{nk}}{ \omega_{nk} } \hat a_n^{\dagger} \hat a_n  \left(\hat b_{nk}^\dagger-\hat b_{nk} \right)  } =\prod_{n} e^{\hat a_n^{\dagger} \hat a_n \tilde S_n}.
\end{align}
In doing so, the Hamiltonian \eqref{eq:generalHamiltonian} is modified $J_{mn}\rightarrow J_{mn}\kappa_{m}\kappa_{n}$ and $H_{\rm sb}\rightarrow \frac 12 \sum_{mn} V_{mn}$, where
\begin{eqnarray}
	\kappa_{m}&=& e^{-\frac 12\sum_{k} \frac{g_{mk}^2 }{\omega_{mk}}   \coth\left( \beta \omega_{mk}/2 \right)}\nonumber \\
	\rightarrow \kappa_m &=& e^{- \int_{0}^{\infty} \frac{d\omega}{2\pi}\mathcal{J}_m(\omega){\omega^2} \coth(\beta_m\omega /2) } , \nonumber \\
	V_{mn} &=& e^{\hat S_m}e^{-\hat S_n}- \kappa_{mn}.
\end{eqnarray}
Later we want to construct the Redfield equation describing the reduced dynamics of the system. For this reason, we need the correlation function
\begin{align}
\left< V_{mn}(t) V_{m'n'} \right>_b  =& \kappa_{m}\kappa_{m'}\kappa_{n}\kappa_{n'}  \\
&\times  \left[ e^{  \lambda_{m'n'}^{m}  \phi_m (t) + \lambda_{n'm'}^{n}  \phi_n (t)} -1 \right] ,
\label{eq:polaronBathCorrelationFkt}
\end{align}
where $\lambda_{m'n'}^{m} \equiv \delta_{m,n'}-\delta_{m,m'} $ determines the sign and
\begin{equation}
\phi_m(t)=\int_{0}^{\infty} \frac{d\omega}{2 \pi } \frac{\mathcal J_m (\omega)} {\omega^2}   \left[ \cos(\omega t ) \coth(\beta_m \omega / 2) + i \sin(\omega t)  \right] .
\end{equation}

\section{Redfield equation in Born-Markov approximation}
\label{sec:masterEquation}

To derive the equation of motion of the reduced density matrix Eq.~\eqref{eq:masterequation}, we follow the steps in~\cite{Schaller2014a}. A general Hamiltonian describing the coupling to the baths reads
\begin{equation}
 H= H_{s} + H_{b} + \sum_{\alpha} \hat B_\alpha \otimes \hat A_\alpha,
\end{equation}
where $H_s$ and $H_{b}$ denote the Hamiltonians of the system and the bath respectively. They are coupled by a sum of products of system ($\hat A_\alpha$) and  bath ($\hat B_\alpha$) operators.

The  Redfield equation in second-order perturbation theory with Born-Markov approximation reads
\begin{eqnarray}
 \frac d{dt}\rho = -i \left[ H_s, \rho \right] - \sum_{\alpha,\beta}\left[\hat A_{\alpha}, G_{\alpha,\beta}\right] ,
\label{eq:reducedLiouvilleEqStationary} 
\end{eqnarray}
where we have defined
\begin{eqnarray}
G_{\alpha,\beta} = \lim_{t\rightarrow\infty}
 \int_{0}^{t}   \mathcal     C_{\alpha,\beta}(\tau)   \hat A_{\beta} (-\tau) \rho\mathcal   
- \mathcal    C_{\beta,\alpha}(-\tau) \rho\hat A_{\beta} (-\tau) d\tau \nonumber
\label{eq:auxFunction}
\end{eqnarray}
with
\begin{equation}
\hat A_{\beta}(\tau) = e^{i H_s\tau  } \hat A_{\beta} e^{-i H_s \tau }  \nonumber,
\end{equation}
and the bath correlation function
\begin{align}
 \mathcal C_{\alpha,\beta}(\tau) &= \text{tr} \left[e^{i H_b \tau }\hat B_\alpha  e^{-i H_b \tau }  B_\beta  \rho_{b}(0) \right]\\
 &= \left<\hat B_\alpha (t)\hat B_\beta \right>_b.
\end{align}
Here $\rho_{b}(0)$ denotes the initial state of the bath.

We continue to evaluate the integral expression in the eigenbasis of $H_s$, which we denote by $\left| a\right>$, which corresponds to the eigenvalues $\omega_a$. Let us for example consider the term
\begin{eqnarray}
\mathcal G_{\alpha,\beta}^{(1)} &=& \lim_{t\rightarrow\infty}  \int_{0}^{t} \mathcal  C_{\alpha,\beta}(\tau) \hat A_{\beta} (-\tau)  \rho \; d\tau \\
&=&  \int_{0}^{\infty} \sum_{abcd} \mathcal  C_{\alpha,\beta}(\tau)  A_{ab}^{\alpha } \rho_{cd } \left|a\right> \left<b  \right|  \left|c\right> \left<d  \right|  
\nonumber e^{-i(\omega_a -\omega_b)\tau }           \nonumber \\
&=&  \int_{0}^{\infty} \mathcal   C_{\alpha,\beta}(\tau) \sum_{abc} A_{ab}^{\alpha } \rho_{bc } \left|a\right> \left<c  \right|  e^{-i(\omega_a -\omega_b)\tau}   , \nonumber
\label{eq:auxFunction}
\end{eqnarray}
where we have used the expansion of the operators
\begin{eqnarray}
\rho = \sum_{cd}  \rho_{cd } \left|c\right> \left<d  \right|, \qquad
\hat A_\alpha = \sum_{ab} A_{ab}^{\alpha } \left|a\right> \left<b  \right|
\end{eqnarray}
Now we use that we consider large times, so that we can set $t\rightarrow \infty$. We further use that
\begin{align}
 \int_{0}^{\infty} e^{-i \omega \tau} \mathcal C_{\alpha,\beta }(\tau)d\tau
= \tilde C_{\alpha, \beta}(\omega) ,
\end{align}
so that we finally find
\begin{eqnarray}
0= - i  \left(\omega_{ad} \right) \rho_{ad} 
&-& \sum_{\alpha,\beta , bc } A^{\alpha  }_{ab}A^{\beta }_{bc}\rho_{cd} \tilde C_{\alpha,\beta}( \omega_{bc})  \nonumber \\
&+& \sum_{\alpha,\beta , bc } A^{\beta  }_{ab} \rho_{bc} A^{\alpha  }_{cd}   \tilde C_{\alpha,\beta}(- \omega_{cd})  \nonumber \\
&+& \sum_{\alpha,\beta , bc } A^{\beta }_{ab}\rho_{bc}A^{\alpha  }_{cd} \tilde C_{\alpha,\beta}( \omega_{ab})\nonumber \\
&-& \sum_{\alpha,\beta , bc } \rho_{ab}A^{\alpha  }_{bc} A^{\beta   }_{cd}   \tilde C_{\alpha,\beta}(- \omega_{bc}) .
\end{eqnarray}

For the polaron-transformed Hamiltonian discussed in Sec.~\ref{sec:polaronTrafo}, we identify
\begin{equation}
\sum_{\alpha} \hat B_\alpha\otimes \hat A_\alpha = \sum_{n=1,2}  J_{n} \hat a_{3}^\dagger \hat a_{n}\otimes \hat V_{3,n}+ \text{ h.c. }
\end{equation}
The bath correlation functions  $\mathcal C_{\alpha,\beta }(\tau) $ of the $\hat V_{3,n}$ are given in Eq.~\eqref{eq:polaronBathCorrelationFkt}.

\section{Approximate current formula}
\label{app:apprFormula}

\label{app:CurrentFormula}

Considering the case of a vanishing system-bath coupling $\gamma=0$, the stationary state of the Redfield equation with Born, Markov and secular approximation fulfills
\begin{align}
 P_{0}  k_{\lambda}^+ &= P_{\lambda}k_{\lambda}^+\nonumber , \\
 P_{0} &= \frac{1}{1+\sum_{\lambda>0} \frac{k_{\lambda}^+}{k_{\lambda}^-}},
\end{align}
with $P_{\lambda}$ denoting the probabilities to be in eigenstate $\left| \lambda \right>$, and $k_{\lambda}^\pm= k_{1,\lambda}^\pm +k_{2,\lambda}^\pm$ defined in Eq.~\eqref{eq:transitionRate}. The current can be expressed as
\begin{align}
	I_{s} &= \sum_{\lambda>0} k_{1,\lambda}^{+} P_{0} -k_{1,\lambda}^{-} P_{\lambda}\\
	&= \sum_{\lambda>0} P_{0} \frac{k_{1,\lambda}^{+}  k_{2,\lambda}^{-} -   k_{1,\lambda}^{-}  k_{2,\lambda}^{+}  }{   k_{1,\lambda}^{-}   +  k_{2,\lambda}^{-}   } \\
	&\equiv \sum_{\lambda>0} I_{s}^{(\lambda)}
	\label{eq:CurrentLambda}
\end{align}

Let us now introduce the Green's function by
\begin{equation}
	G(z)\equiv   i \frac{1} {   z+ i  H_s} = i \sum_{\lambda} \frac{\left|\lambda \right>\left< \lambda \right|}{ z+ i E_{\lambda} }
	\label{eq:defGreensFunction}
\end{equation}
for a complex-valued $z$. Here $E_\lambda$ and $\left|\lambda \right>$ denote the eigenvalues and eigenstates of $H_{\rm s}$. For a real $E$, we also define the retarded (r) and advanced (a) Green's function by $G^{r,a}(E)=\lim_{\delta\rightarrow 0^+}\equiv G (- i E\pm  \delta)$. Using the Dirac identity
\begin{equation}
	\lim_{\delta\rightarrow 0^+} \frac 1{x+ i\delta} = \mathcal P (x) + i \pi \delta(x)
\end{equation}
it is not hard to see that
\begin{equation}
\frac 1{2i} \left<n \right| G^{r}(E_{\lambda})-G^{a}(E_{\lambda})  \left| m \right>  = \left<n \right.\left| \lambda \right>  \left<\lambda \right.\left| m \right>  \equiv G^{\lambda,-}_{nm}.
\end{equation}
For this reason, we can rewrite the expression for the current as
\begin{align}
	k_{n,\lambda}^{+}  k_{m,\lambda}^{-} &= \left| G^{\lambda,-}_{\lambda,nm}\right|^2 \Gamma_{n}(E_\lambda) \Gamma_{m}(E_\lambda) f_n(E_\lambda) \left[1- f_m(E_\lambda) \right]\nonumber \\
	k_{n,\lambda}^{+} &= G^{\lambda,-}_{\lambda,nn}\cdot \Gamma_{n}(E_\lambda) f_n(E_\lambda)\nonumber \\
	k_{n,\lambda}^{-} &= G^{\lambda,-}_{\lambda,nn}\cdot \Gamma_{n}(E_\lambda)\left[1- f_n(E_\lambda) \right]
\end{align}
The Green's function can be easily obtained by inversion of a $3\times3$ function according to Eq.~\eqref{eq:defGreensFunction}. In doing so we find
\begin{equation}
 \left<n \right| G(z) \left| m \right>= i \frac{  \left[g(z)\right]_{nm} }{\mathcal G(z)} ,
\end{equation}
where
\begin{eqnarray}
 \left[g(z)\right]_{11} &=&  (z+i \epsilon_2 )(z+ i \epsilon_3 ) + J_2 J_2 \nonumber , \\
 \left[g(z)\right]_{21} &=&  -i J_1 e^{-i \phi} (z + i \epsilon_3 ) - J_2 J_3 \nonumber ,  \\
\left[g(z)\right]_{31} &=&  - J_1 e^{-i \phi} J_2 -  i J_3 (z+ i \epsilon_2 )\nonumber , \\
\left[g(z)\right]_{12} &=&  -J_2 J_3 -  i J_1 e^{i \phi}(z+ i \epsilon_3 ) \nonumber,  \\
 \left[g(z)\right]_{22} &=&  (z+ i \epsilon_1 )(z+ i \epsilon_3 ) + J_3 J_3\nonumber ,  \\
 \left[g(z)\right]_{32} &=&  -(z+ i \epsilon_1 )i J_2 -J_1 e^{i \phi} J_3 \nonumber,  \\
 \left[g(z)\right]_{13} &=&   - J_1 J_2 e^{i \phi} - i J_3 (z + i \epsilon_2 )\nonumber , \\
 \left[g(z)\right]_{23} &=&  - i (z+i \epsilon_1 )J_2 -J_1 e^{-i \phi} J_3\nonumber , \\
 \left[g(z)\right]_{33} &=&  (z+ i \epsilon_1 )(z  + i  \epsilon_2 ) + J_1 J_1, \nonumber 
\end{eqnarray}
and
\begin{align}
\mathcal G(z) &\equiv \det(z + i H_e) \nonumber \\
&=(z+ i \epsilon_1)(z+ i \epsilon_2)(z+ i \epsilon_3) - 2 i J_1 J_2 J_3 \cos(\phi) \nonumber \\
&+ (z+ i \epsilon_2) J_3 ^2 + J_1^2 (z+ i \epsilon_3) +  J_2^2 (z+ i \epsilon_1).\\
\nonumber
\end{align}
For a notational reason, we have introduced $J_n = J_{n,n+1}^{(0)}$ for $n=1,2,3$ with $n=4$ corresponds to $n=1$.
Putting now all ingredients together, we obtain
\begin{equation}
 I_{s}^{(\lambda)} =  \mathcal N_\lambda \left|J_{1} \cdot  (E_{\lambda}- \epsilon_3 ) +  J_{2} J_{3} e^{ \phi} \right|^2,
\end{equation}
 with 
 \begin{align}
 \mathcal N _\lambda &= \frac {P_{0} }{\mathcal K_\lambda}
\frac{\Gamma_{n} (E_{\lambda}) \Gamma_{m} (E_{\lambda})    f_{n} (E_\lambda) \left[ 1-f_{n} (E_\lambda) \right]} 
 {\sum_{n=1,2}\Gamma_n (E_{\lambda}) g_{nn}^{\lambda} \left[ 1- f_n(E_\lambda)\right]} ,\nonumber \\
 P_{0}&= \frac 1 { 1+ \sum_{\lambda>0}\frac{\sum_{n=1,2} \Gamma_n (E_{\lambda}) g_{nn}^{\lambda } f_n(E_\lambda) }{\sum_{n=1,2} \Gamma_n (E_{\lambda}) g_{nn}^{\lambda } \left[ 1- f_n(E_\lambda)\right]  }} \nonumber \\
 \mathcal K_\lambda &=\prod_{\lambda'\neq0,\lambda} (-i) (E_{\lambda}-E_{\lambda'})
 \nonumber
 \end{align}
 where we have defined $g_{nm}^{\lambda} \equiv \left[ g(-i E_\lambda ) \right]_{n,m}   $.

\section{Path interference}
\label{app:pathInterference}

We are interested into the transition probability from site $n=1$ to site $n=2$ as discussed in Sec.~\ref{sec:pathInterference}. To this end, we have to calculate the matrix element $U^{(0)}_{1,2}(t)$ and $U^{(1)}_{1,2}(t)$ of the (conditioned) time evolution operators. 
For simplicity we assume $\epsilon_{1,2}=0$ in the following.

To obtain $U^{(0,1)}_{1,2}(t)$,
we use that the time-evolution operator is related to the Green's function defined in Eq.~\eqref{eq:defGreensFunction} by a Laplace transformation. In doing so, we obtain
\begin{equation}
\int_{0}^{\infty} e^{- z t } e^{- i H_{s} t} = \frac{1}{z+ i H_{s}} = - i G(z).
\end{equation}
Setting $J_{2}=J_{3}=0$ and $J_1=J$, we  apply the inverse Laplace transformation to 
\begin{equation}
	- i \left<2 \right| G(z) \left| 1 \right> =\frac{ -i J e^{-i \phi}  }{  z^2 + J ^2 } \rightarrow  i e^{i \phi} \sin(J t)=U^{(0)}_{1,2}(t)
\end{equation}
to obtain $U^{(0)}_{1,2}(t)$. To calculate $U^{(1)}_{1,3}(t)$, we set $J_{2}=J_{3}=J$  and $J_1=0$.  Applying  the inverse Laplace transformation to
\begin{align}
	- i  \left<2 \right| G(z) \left| 1 \right>&=\frac{ - J^2  }{  z (z+ i \epsilon_3)+ z J + z J } \\
	&=\frac{ - J^2  }{  z (z-z_1)(z-z_2) }
\end{align}
with 
\begin{equation}
z_{1,2} = -i\frac{\epsilon_3}{2}\pm i \sqrt{\frac {\epsilon_3^2}{4} + 2 J^2 },
\end{equation}
 we obtain
\begin{equation}
 U^{(1)}_{1,2} = - \frac{J^2}{z_1 z_2} - \frac{J^2 e^{z_1 t} }{z_1 (z_1-z_2)  }  - \frac{J^2 e^{z_2 t} }{z_1 (z_2-z_1)  },
\end{equation}
and if additionally assuming $\epsilon_3=0$, we find
\begin{equation}
 U^{(1)}_{1,2} (t) = -\frac 12 + \frac 12 \cos(\sqrt{2} J t   ).
\end{equation}

\end{document}